\documentclass[12pt]{article}
\usepackage{amssymb,amsmath,epsfig}
\allowdisplaybreaks
\begin{document}

\title{\bf Instability Analysis of Cylindrical Stellar Object in Brans-Dicke Gravity}

\author{M. Sharif $^1$ \thanks{msharif.math@pu.edu.pk} and Rubab
Manzoor $^2$
\thanks{rubab.manzoor@umt.edu.pk}\\
$^1$ Department of Mathematics, University of the Punjab,\\
Quaid-e-Azam Campus, Lahore-54590, Pakistan.
\\$^2$ Department of Mathematics,\\ University of Management and
Technology,\\
Johar Town Campus, Lahore-54782, Pakistan.}
\date{}
\maketitle

\begin{abstract}
This paper investigates instability ranges of a cylindrically
symmetric collapsing cosmic filamentary structure in Brans-Dicke
theory of gravity. For this purpose, we use perturbation approach in
the modified field equations as well as dynamical equations and
construct a collapse equation. The collapse equation with adiabatic
index ($\Gamma$) is used to explore the instability ranges of both
isotropic as well as anisotropic fluid in Newtonian and
post-Newtonian approximations. It turns out that the instability
ranges depend on the dynamical variables of collapsing filaments. We
conclude that the system always remains unstable for $0<\Gamma<1$
while $\Gamma>1$ provides instability only for the special case.
\end{abstract}

\section{Introduction}

Dark energy and gravitational collapse are the most fascinated and
interesting phenomena of cosmology as well as gravitational physics.
Number of astronomical observations  such as Supernova type I, Sloan
Digital Sky Survey, large scale-structure, Wilkinson Microwave
Anisotropy Probe, galactic cluster emission of X-rays and weak
lensing describe accelerated behavior of the expanding universe
\cite{1}. It is suggested that a mysterious type of energy known as
dark energy is responsible for this accelerated expansion of the
universe. This induces the problem of correct theory of gravity and
thus numbers of modified theories of gravity are constructed using
modified Einstein-Hilbert actions. Brans-Dicke (BD) theory is one of
the most explored examples among various modified theories that
provides convenient evidences of various cosmic problems  like
inflation, early and late behavior of the universe, coincidence
problem and cosmic acceleration \cite{2}. This is a generalized form
of general relativity (GR) which is constructed by the coupling of
scalar field $\phi$ and tensor field $R$. It contains a constant
coupling parameter $\omega_{BD}$ (tuneable parameter) which can be
adjusted according to suitable observations. This theory is
compatible with Mach's principle, weak equivalence principle and
Dirac's large number hypothesis \cite{3}. It is also consistent with
solar system observations and experiments (weak field regimes test)
for $|\omega|\geq40,000$ \cite{5}.

Gravitational collapse is a process in which stable stellar objects
turn into unstable ones under the effects of their own gravity. The
formation and dynamics of large scale structures such as stars,
celestial cluster and galaxies are investigated through this
phenomenon. It is believed that different instability ranges for
astronomical bodies lead to different structure formation of
collapsing models. Chandrasekhar \cite{6} was the first who explored
stability ranges of a spherically symmetric isotropic fluid in GR.
He used equation of state involving adiabatic index $(\Gamma)$ and
concluded that the fluid remains unstable for $\Gamma<\frac{4}{3}$.
Later on, many researchers \cite{7, 7'} investigated dynamical
instability of different types of fluids (anisotropic fluid,
adiabatic, non adiabatic as well as shearing viscous fluid) in
spherical as well as cylindrical configurations and found that
stability ranges depend on physical properties of the respective
fluid.

It is believed that the study of collapse phenomenon in modified
theories may reveal modification hidden in the formation of
astronomical structures \cite{8}. In 1969, Nutku \cite{9} explored
instability ranges of spherically symmetric isotropic fluid in BD
theory and concluded that BD fluid remains unstable for
$\Gamma>\frac{4}{3}$. Kwon et al. \cite{10} discussed instability
analysis of the Schwarzschild black hole in BD gravity. Sharif and
Kauser \cite{11} investigated stability ranges for spherical as well
as cylindrical collapsing models in $f(R)$ theory and found that
instability ranges depend upon characteristics of fluids and dark
energy components. Sharif and Yousaf \cite{12} studied the effects
of electromagnetic field on instability ranges for various models of
$f(R)$ gravity. Sharif and Rani \cite{13} explored dynamical
instability of spherically symmetric fluid in $f(T)$ theory and
concluded that modified terms control instability ranges. In a
recent paper \cite{14}, we have discussed collapse of spherically
symmetric anisotropic BD fluid through instability analysis and
found that $0<\Gamma<1$ always leads to unstable configuration while
$\Gamma>1$ provides instability only for one particular case.

The behavior of filamentary structures has important implications
for the formation of structure in the universe. Galaxy filaments are
the largest known cosmic structures in the universe. The filamentary
structure is always present in the interstellar medium and
instabilities within these filaments create dense medium (dense
core) where stars form \cite{14a}. N-body simulations of formation
of large scale structure describe a wide range of filaments (with a
cluster of galaxies forming at the intersection of filaments)
\cite{14b}. Filamentary structures are associated with cosmic web on
the large scales and are used to describe tidal tails (thrown off by
merging galaxies) on the small scales \cite{14b'}. In order to
understand fragmentation of filament structures, cylindrically
symmetric filament models are widely studied \cite{14c}.

In this paper, we investigate dynamical instability of cylindrically
symmetric filaments collapsing structure in BD gravity. The paper is
organized in the following format. The next section discusses BD
equations, Darmois junction conditions and dynamical equations. In
section \textbf{3}, we use perturbation technique to construct
hydrostatic equilibrium (collapse equation) and describe instability
ranges (at Newtonian and post-Newtonian (pN) limits) for isotropic
as well as an anisotropic fluid distributions. Finally, the last
section summarizes the results.

\section{Brans-Dicke Theory and Dynamical Equations}

The BD theory (with self-interacting potential $V(\phi)$ ) has the
following action \cite{3}
\begin{equation}\label{1}
S=\int d^{4}x\sqrt{-g} [\phi
R-\frac{\omega_{BD}}{\phi}\nabla^{\mu}{\phi}\nabla_{\mu}{\phi}-V(\phi)+\emph{L}_{m}],
\end{equation}
where $8\pi G_0=c=1$ and $\emph{L}_{m}$ represents matter
distribution. Varying Eq.(\ref{1}) by $g_{\alpha\beta}$ and $\phi$,
we obtain the following BD equations
\begin{eqnarray}\label{2}
G_{\alpha\beta}&=&\frac{1}{\phi}(T_{\alpha\beta}^{m}+T_{\alpha\beta}^{\phi}),\\\label{3}
\Box\phi&=&\frac{T^{m}}{3+2\omega_{BD}}
+\frac{1}{3+2\omega_{BD}}[\phi\frac{dV(\phi)}{d\phi}-2V(\phi)].
\end{eqnarray}
Here $G_{\alpha\beta}$ is the Einstein tensor, $T_{\alpha\beta}^{m}$
is the energy-momentum tensor for matter distribution with $T^{m}$
as its trace and $\square$ represents d'Alembertian operator. The
energy distribution due to scalar field is given by
\begin{equation}\label{4}
T^{\phi}_{\alpha\beta}=\phi_{,\alpha;\beta}
-g_{\alpha\beta}\Box\phi+\frac{\omega_{BD}}{\phi}[\phi_{,\alpha}\phi_{,\beta}
-\frac{1}{2}g_{\alpha\beta}\phi_{,\mu}\phi^{,\mu}]
-\frac{V(\phi)}{2}g_{\alpha\beta}.
\end{equation}
Equation (\ref{2}) gives the BD field equations and (\ref{3}) is a
wave equation for the evolution of scalar field.

We split $4$D geometry into interior and exterior regions by
considering a timelike 3D hypersurface $\Sigma^{(e)}$ as an external
boundary of the respective cylindrical body \cite{7'}. The interior
region of a collapsing cylindrical filamentary structure is
represented by
\begin{equation}\label{5}
ds^2_-=A^2(t,r)dt^{2}-B^2(t,r)dr^{2}-C^2(t,r)d\phi^{2}-dz^{2},
\end{equation}
where we consider comoving coordinates inside the hypersurface. In
order to preserve cylindrical symmetry, the coordinates satisfy the
following constraints
\begin{eqnarray}\nonumber
-\infty\leq t\leq\infty,\quad0\leq r<\infty, \quad
-\infty<z<\infty,\quad0\leq\phi\leq2\pi.
\end{eqnarray}
In stationary or static region, a scalar field becomes constant and
all stationary black holes in BD gravity are identical with GR
solutions \cite{15}. Therefore, for exterior region to
$\Sigma^{(e)}$, we take line element of static cylindrical black
hole given by
\begin{equation}\label{6}
ds^2_+=-\frac{2M}{r}d\nu^2+2drd\nu-r^2(d\phi^2+\gamma^2dz^2),
\end{equation}
where $M,~\nu$ and $\gamma$ describe the total gravitating mass,
retarded time, and arbitrary constant, respectively \cite{16}. The
interior region is filled with anisotropic matter distribution
represented by
\begin{equation}\label{7}
T_{\alpha\beta}^{m}=(\rho+p_{r})u_{\alpha}u_{\beta}-p_{r}g_{\alpha\beta}
+(p_{z}-p_{r})S_{\alpha}
S_{\beta}+(p_{\phi}-p_{r})K_{\alpha}K_{\beta},
\end{equation}
where $\rho,~p_{r},~p_{\phi}$ and $p_{z}$ indicate energy density
and principal pressure stresses, respectively. The four velocity
$u_{\alpha}$, unit four-vectors $S_{\alpha}$ and $K_{\alpha}$ are
calculated as $u_{\alpha}=A\delta_{\alpha}^{0},~
S_{\alpha}=\delta_{\alpha}^{3}$ and
$K_{\alpha}=C\delta_{\alpha}^{2}$ satisfying
$u^{\alpha}u_{\alpha}=1,~S^{\alpha}S_{\alpha}=K^{\alpha}K_{\alpha}=-1,~
S^{\alpha}u_{\alpha}=K^{\alpha}u_{\alpha}=S^{\alpha}K_{\alpha}=0$.
For the interior region, the BD equations are given in Appendix
\textbf{A}.

Junction conditions provide smooth connection between interior and
exterior regions over $\Sigma^{(e)}$. We consider Darmois junction
conditions to discuss connection between two regions \cite{7'} and
for this purpose we take C-energy (mass function) \cite{17} given by
\begin{equation}\label{14}
\tilde{E}(t,r)=m(t,r)=\frac{1}{8}(1-l^{-2}\nabla^\beta
\tilde{r}\nabla_{\beta}\tilde{r}).
\end{equation}
Here $\tilde{E}(t,r)$ is the gravitational energy per unit specific
length of the cylinder, $\tilde{r}$ represents the areal radius,
$\mu$ shows the circumference radius and $l$ indicates specific
length. These are given as follows
\begin{equation}\nonumber
\tilde{r}=\mu l,\quad \mu^2=\xi_{(1)\beta}\xi^{\beta}_{(1)},\quad
l^2=\xi_{(2)\beta}\xi^{\beta}_{(2)},
\end{equation}
where $\xi_{(1)}=\frac{\partial}{\partial\theta}$ and
$\xi_{(2)}=\frac{\partial}{\partial z}$ are the respective Killing
vectors. For the interior spacetime, Eq.(\ref{14}) takes the form
\begin{equation}\label{15}
m(t,r)=\frac{l}{8}\left(1+\frac{\dot{C}^2}{A^2}
-\frac{C'^2}{B^2}\right),
\end{equation}
where dot and prime show derivatives with respect to $t$ and $r$,
respectively. Since in BD gravity, scalar field and metric tensor
are indicated as gravitational variables, therefore
$\phi=\phi_{\Sigma^{(e)}}=constant$ at the hypersurface
$\Sigma^{(e)}$. The continuity of first and second fundamental forms
(Darmois conditions) yield the following relations
\begin{eqnarray}\nonumber
&&r=r_{\Sigma^{(e)}}=constant,\quad
m(t,r)-M\overset{\Sigma^{(e)}}{=}\frac{l}{8},\quad
l\overset{\Sigma^{(e)}}{=}4C,\\\label{15*}
&&\frac{p_r}{\phi}\overset{\Sigma^{(e)}}{=}\frac{-T^{\phi}_{11}}{B^2}
-\frac{T^{\phi}_{01}}{AB}=-\frac{V(\phi)}{2\phi}.
\end{eqnarray}
Dynamical equations obtained from the contracted Bianchi identities
describe the conservation of total energy of the system given by
\begin{eqnarray}\label{16}
\left(\frac{T^{\alpha\beta}_{m}}{\phi}+\frac{T^{\alpha\beta}_{\phi}}{\phi}\right)_{;\alpha}u_{\beta}=0,\quad
\left(\frac{T^{\alpha\beta}_{m}}{\phi}+\frac{T^{\alpha\beta}_{\phi}}{\phi}\right)_{;\alpha}
\chi_{\beta}=0,
\end{eqnarray}
where $\chi_{\beta}=-B\delta^{1}_{\beta}$ (unit four-vector) which
provides
\begin{eqnarray}\label{17}
&&\left[\frac{\dot{\rho}}{A}
-\frac{\rho\dot{\phi}}{\phi^2A}+(\rho+p_r)\frac{\dot{B}}{AB}+(\rho+p_{\phi})
\frac{\dot{C}}{AC}\right]+ K_{1}=0,\\\label{18}
&&\left[\frac{p_r'}{B}+\frac{\phi'p_{r}}{\phi^{2}B}
+(\rho+p_r)\frac{A'}{AB}+(p_r-p_{\phi})\frac{C'}{BC}\right]+K_{2}=0,
\end{eqnarray}
$K_{1}$ and $K_{2}$ are mentioned in Appendix \textbf{A}.

\section{Instability Analysis}

Here, we use perturbation approach to construct collapse equation
which will be used for instability analysis. We assume that
initially, the system is in static equilibrium (metric as well as
material parts have radial dependence only) and after that all the
dynamical variables along with metric functions are perturbed and
time dependence appears \cite{7'}. The scalar field, scalar
potential and metric tensors have the same time dependence, while
the density and pressure bear the same time dependence as follows
\begin{eqnarray}\label{19}
A(t,r)&=&A_0(r)+\epsilon T(t)a(r),\\\label{20}
B(t,r)&=&B_0(r)+\epsilon T(t)b(r),\\\label{21}
C(t,r)&=&C_0(r)+\epsilon T(t)c(r),\\\label{22}
\phi(r,t)&=&\phi_{o}(r)+\epsilon T(t)\Phi(r),\\\label{23}
p_r(t,r)&=&p_{r0}(r)+\epsilon {\bar{p}_r}(t,r),\\\label{24}
p_{\phi}(t,r)&=&p_{\phi0}(r)+\epsilon{\bar{p}_\phi}(t,r),\\\label{25}
\rho(t,r)&=&\rho_0(r)+\epsilon{\bar{\rho}}(t,r),\\\label{26}
V(\phi)&=&V_0(r)+\epsilon T(t){\bar{V}(r)},
\end{eqnarray}
where $0<\epsilon\ll1$ and the static distribution is expressed by
zero subscript. For static and perturbed configurations of the field
as well as dynamical equations, we take $C_0=r$. The static
configuration of BD formalism, perturbed form of BD equations and
junction condition (\ref{15*}) are given in Appendix \textbf{A}.

The perturbed distribution of first Bianchi identity gives
\begin{eqnarray}\label{40}
\bar{\rho}&=&-\left[\frac{(\rho_0+p_{r0})b}{B_0}
+\frac{(\rho_0+p_{\phi0})c}{r}+\frac{\Phi\rho_{0}}{\phi_{0}}
+A_{0}\phi_0\bar{K_{1}}\right]T.
\end{eqnarray}
The perturbed form of Eq.(\ref{18}) provides
\begin{eqnarray}\label{41}
\bar{p}_{r}'+(\bar{\rho}+\bar{p}_{r})\frac{A_{0}'}{A_0}
+(\bar{p}_{r}-\bar{p}_{\phi})\frac{1}{r}+\frac{\bar{p}_{r}\phi_{0}'}
{\phi_{0}B_{0}}+\bar{K}_{2}\phi_{0}B_{0}=0,
\end{eqnarray}
where $\bar{K}_{1}$ and $\bar{K}_{2}$ are given in appendix
\textbf{A}. Equation (\ref{35}) along with junction conditions
provides
\begin{equation}\label{44}
T(t)=c_1\exp({\gamma_{\Sigma^{(e)}}t})+
c_2\exp({\lambda_{\Sigma^{(e)}}t}),
\end{equation}
where $\gamma_{\Sigma^{(e)}}=+\sqrt{\frac{v}{u}}$,
$\lambda_{\Sigma^{(e)}}=-\sqrt{\frac{v}{u}}$ with
\begin{eqnarray}\nonumber
u\overset{\Sigma^{(e)}}{=}\frac{\Phi}{\phi_{0}A^{2}_0}
-\frac{2c}{rA^2_{0}} ,\quad
v\overset{\Sigma^{(e)}}{=}\frac{\Phi}{\phi_{0}}
\left[\frac{\omega_{BD}}{\phi_{0}}-\frac{1}{B_{0}r}\right]
\end{eqnarray}
and $c_1,~c_2$ indicate arbitrary constants. Equation (\ref{44})
shows static and non-static distributions leading to stable as well
as unstable phases of gravitating system. For instability analysis,
we assume that when the instability phase begins, the system was in
complete hydrostatic equilibrium, i.e., ($t=-\infty,~T(-\infty)=0$).
Using this assumption in Eq.(\ref{44}), we have $c_2=0$ whereas
$c_1=-1$ is chosen arbitrarily. The corresponding result is
described by
\begin{eqnarray}\label{45}
T(t)=-\exp({\gamma_{\Sigma^{(e)}}t}).
\end{eqnarray}
For a real instability regime, we assume only positive values of
$\frac{v}{u}$.

For the investigation of instability ranges, we use an equation of
state involving adiabatic index $\Gamma$ \cite{18} given by
\begin{equation}\label{46}
{\bar{p}_j}=\Gamma \frac{p_{j0}}{\rho_0+p_{j0}}\bar{\rho}.
\end{equation}
The adiabatic index evaluates variation of principal stresses
(pressures) with respect to density and represents rigidity of the
gravitating fluid. We consider $\Gamma$ to be constant throughout
the stability analysis of the fluid. Equations (\ref{40}) and
(\ref{46}) lead to
\begin{eqnarray}\label{47}
\bar{p}_r&=&-\Gamma
\left[\frac{b}{B_0}p_{r0}+\frac{c}{r}\frac{\rho_0+p_{\phi0}}{\rho_0
+p_{r0}}p_{r0}+\frac{p_{r0}}{\rho_0+p_{r0}}\frac{\Phi
\rho_{0}}{\phi_{0}}+\frac{p_{r0}A_{0}
\phi_0}{\rho_0+p_{r0}}\bar{K_{1}}\right]T,\\\label{48*}
\bar{p}_\phi&=&-\Gamma\left[\frac{b}{B_0}\frac{\rho_0+p_{r0}}{\rho_0
+p_{\phi0}}p_{\phi0}+\frac{c p_{\phi0}}{r} +\frac{p_{\phi0}\Phi
\rho_{0}}{(\rho_0+p_{\phi0})\phi_{0}}
+\frac{p_{\phi0}A_{0}\phi_0}{\rho_0+p_{\phi0}}\bar{K_{1}} \right]T.
\end{eqnarray}
Using Eqs.(\ref{31}), (\ref{40}), (\ref{47}) and (\ref{48*}) in
(\ref{41}), we construct a hydrostatic equation given by
\begin{eqnarray}\nonumber
&&\Gamma\left[p_{ro}\left[\frac{bT}{B_{0}}
+\frac{(\rho_{0}+p_{\phi0})}{(\rho_{0}+p_{r0})}\frac{cT}{r}
+\frac{1}{\rho_{0}+p_{ro}}A_{0}\phi_{0}\bar{K}_{1}T\right]\right]'
-\frac{\Gamma}{r}\left[p_{ro}\left[\frac{bT}{B_{0}}\right.\right.\\\nonumber
&&\left.\left.+\frac{(\rho_{0}+p_{\phi0})}{(\rho_{0}+p_{r0})}\frac{cT}{r}
+\frac{p_{r0}}{\rho_{0}+p_{ro}}A_{0}\phi_{0}\bar{K}_{1}T\right]\right]
+\frac{p_{\phi0}}{r}\Gamma\left[p_{\phi o}\left[\frac{bT}{B_{0}}
\frac{(\rho_{0}+p_{r0})}{(\rho_{0}+p_{\phi0})}\right.\right.\\\nonumber
&&\left.\left.+\frac{cT}{r}+\frac{1}{\rho_{0}+p_{\phi
o}}A_{0}\phi_{0}\bar{K}_{1}T\right]\right]-
\Gamma\left[p_{ro}\left[\frac{bT}{B_{0}}
+\frac{(\rho_{0}+p_{\phi0})}{(\rho_{0}+p_{r0})}\frac{cT}{r}
+\frac{1}{\rho_{0}+p_{ro}}\right.\right.\\\nonumber
&&\left.\left.\times A_{0}\phi_{0}\bar{K}_{1}T\right]\right]
\frac{A'_{0}}{A_{0}}
-\left[\frac{bT}{B_{0}}(\rho_{0}+p_{r0})+(\rho_{0}+p_{\phi0})\frac{cT}{r}
+A_{0}\phi_{0}\bar{K}_{1}T\right]\\\nonumber
&&\times\frac{A'_{0}}{A_{0}}-
\Gamma\left[p_{ro}\left[\frac{bT}{B_{0}}
+\frac{(\rho_{0}+p_{\phi0})}{(\rho_{0}+p_{r0})}\frac{cT}{r}
+\frac{1}{\rho_{0}+p_{ro}}A_{0}\phi_{0}\bar{K}_{1}T\right]\right]
\frac{\phi'_{0}}{\phi_{0}B_{0}}\\\label{48}
&&+\phi_{0}B_{0}\bar{K}_{2}=0.
\end{eqnarray}
This represents the general form of collapse equation which
describes the instability of hydrostatic equilibrium of gravitating
filaments in BD gravity.

\subsection{Isotropic Fluid}

Here, we analyze instability ranges of isotropic fluid in Newtonian
and pN limits. In isotropic fluid, all principal stresses are equal
($p_{r}=p_{\phi}=p_{z}$). Using this condition in Eq.(\ref{48}) we
obtain the corresponding collapse equation
\begin{eqnarray}\nonumber
&&\Gamma\left[p_{ro}\left[\frac{bT}{B_{0}} +\frac{cT}{r}
+\frac{1}{\rho_{0}+p_{ro}}A_{0}\phi_{0}\bar{K}_{1}T\right]\right]'-
\Gamma\left[p_{ro}\left[\frac{bT}{B_{0}}
+\frac{cT}{r}\right.\right.\\\nonumber
&&\left.\left.
+\frac{1}{\rho_{0}+p_{ro}}A_{0}\phi_{0}\bar{K}_{1}T\right]\right]
\frac{A'_{0}}{A_{0}}
-\left[(\rho_{0}+p_{r0})(\frac{bT}{B_{0}}+\frac{cT}{r})
\right.\\\nonumber
&&\left.+A_{0}\phi_{0}\bar{K}_{1}T\right]\frac{A'_{0}}{A_{0}}-
\Gamma\left[p_{ro}\left[\frac{bT}{B_{0}} +\frac{cT}{r}
+\frac{1}{\rho_{0}+p_{ro}}A_{0}\phi_{0}\bar{K}_{1}T\right]\right]
\frac{\phi'_{0}}{\phi_{0}B_{0}}\\\label{49}
&&+\phi_{0}B_{0}\bar{K}_{2}=0.
\end{eqnarray}

\subsection*{Newtonian Limit}

The Newtonian limit in BD theory leads to the following
\begin{eqnarray}\nonumber
&&\rho_0\gg p_{r0},\quad \rho_0\gg p_{\phi0},\quad B_{0}=1,\quad
A_0=1-\frac{m_0}{rc^2},\label{a}\\
&&\phi_0=constant,\quad V_0=\bar{V}=0.
\end{eqnarray}
Using these limits along with (\ref{45}), the collapse condition
turns out to be
\begin{eqnarray}\nonumber
\Gamma\left[(p_{r0}Z_N),_1-\frac{m_{0}}{r^2c^2}p_{r0}Z_N\right]-\rho_{0}Z_{N}\frac{m_{0}}{r^2c^2}
+K_{3}<0,
\end{eqnarray}
which gives
\begin{eqnarray}\label{51}
\Gamma<\frac{\rho_{0}Z_{N}\frac{m_{0}}{r^2c^2}
-K_{3}}{\left[(p_{r0}Z_N),_r-\frac{m_{0}}{r^2c^2}p_{r0}Z_N\right]}.
\end{eqnarray}
Here
\begin{eqnarray}\nonumber
Z_N=\left(b+\frac{c}{r}\right),\quad
K_{3}=-\left[-\frac{a'm_{0}}{r^2c^2\phi_{0}}+
\frac{\Phi}{\phi_{0}}(1-\frac{m_{0}}{rc^2})-2p_{r0}\frac{\Phi}{\phi_{0}}\right].
\end{eqnarray}
This shows that the adiabatic index depends on dynamical properties
such as density, pressure, scalar field. To preserve difference
between configurations of pressure gradient and gravitational
forces, we assume $\Gamma>0$. Thus the celestial objects remain
unstable until (\ref{51}) is satisfied which leads to
\begin{equation}\label{51*}
\frac{\rho_{0}Z_{N}\frac{m_{0}}{r^2c^2}
-K_{3}}{\left[(p_{r0}Z_N),_r-\frac{m_{0}}{r^2c^2}p_{r0}Z_N\right]}>\Gamma>0.
\end{equation}
This leads to the following possibilities:
\begin{enumerate}
\item $\rho_{0}Z_{N}\frac{m_{0}}{r^2c^2}
-K_{3}=\left[(p_{r0}Z_N),_r-\frac{m_{0}}{r^2c^2}p_{r0}Z_N\right]$;
\item
$\rho_{0}Z_{N}\frac{m_{0}}{r^2c^2}
-K_{3}<\left[(p_{r0}Z_N),_r-\frac{m_{0}}{r^2c^2}p_{r0}Z_N\right]$;
\item
$\rho_{0}Z_{N}\frac{m_{0}}{r^2c^2}
-K_{3}>\left[(p_{r0}Z_N),_r-\frac{m_{0}}{r^2c^2}p_{r0}Z_N\right]$.
\end{enumerate}

The first and second case along with (\ref{51*}) show that the
isotropic system becomes unstable for $0<\Gamma<1$. The
corresponding expressions lead to
\begin{eqnarray}\label{53}
p_{r0}=Z^{-1}_N\int^{r}_{r_{0}}Z_N
\left((\rho_{0}-1)\frac{m_{0}}{r^2c^2}-Z^{-1}_{N}K_{3}\right)dr',\\\label{54}
p_{r0}<Z^{-1}_N\int^{r}_{r_{0}}Z_N
\left((\rho_{0}-1)\frac{m_{0}}{r^2c^2}-Z^{-1}_{N}K_{3}\right)dr'.
\end{eqnarray}
These are the constraint expressions for a collapsing cylindrical
isotropic filamentary structure with $0<\Gamma<1$. In the third
case, the denominator is less than its numerator and hence in
(\ref{51*}), $\Gamma$ can be taken greater than $1$. The
corresponding instability constraint is given by
\begin{equation}\label{55}
p_{r0}>Z^{-1}_N\int^{r}_{r_{0}}Z_N
\left((\rho_{0}-1)\frac{m_{0}}{r^2c^2}-Z^{-1}_{N}K_{3}\right)dr'.
\end{equation}
for which $\Gamma>1$ and isotropic cylindrical system becomes
unstable. It is obvious that if the system is unstable for
$\Gamma>1$, then it will also be unstable for $0<\Gamma<1$.

\subsection*{Post-Newtonian Limit}

The pN regimes are found upto order  $c^{-4}$ by taking
\begin{eqnarray}\label{56}
A_0=1-\frac{m_0}{rc^2}+\frac{m_0^2}{r^2c^4},\quad B_0=1+\frac{\alpha
m_0}{rc^2},\quad\phi_0=constant,\quad V_0=\bar{V}=0,
\end{eqnarray}
where
\begin{equation}\nonumber
\alpha=\frac{1+\omega_{BD}}{2+\omega_{BD}}.
\end{equation}
Using pN limits along with Eq.(\ref{45}) in (\ref{48}), we obtain
\begin{eqnarray}\label{58}
0<\Gamma<\frac{\left[(\rho_{0}+p_{r0})X_{pN}\right]
(\frac{m_{0}}{r^2c^2}-2\frac{m_{0}}{r^3c^4})-K_{5}}{\left[\left[p_{ro}X_{pN}\right]'-
\left[p_{ro}X_{pN}\right]
(\frac{m_{0}}{r^2c^2}-2\frac{m_{0}}{r^3c^4})\right]},
\end{eqnarray}
where
\begin{equation}\nonumber
X_{pN}=\left[(b(1-\frac{\alpha m_0}{rc^2})+\frac{c}{r})
+\frac{K_{4}}{\rho_{0}+p_{ro}}\right],
\end{equation}
$K_{4}$ and $K_{5}$ are given in Appendix \textbf{A}. The expression
(\ref{58}) describes condition for instability of a cylindrical
filamentary structure in pN limits. Similar to the Newtonian case,
the system collapses for $0<\Gamma<1$ with the following constraints
\begin{enumerate}
\item $p_{r0}=X^{-1}_{pN}e^{2r(\frac{m_{0}}{r^2c^2}
-\frac{2m^2_{0}}{r^3c^4})+\int^{r}_{r_{0}}Y_{pN}dr'}\int^{r}_{r_{0}}
X_{pN}e^{-2r(\frac{m_{0}}{r^2c^2}
-\frac{2m^2_{0}}{r^3c^4})-\int^{r}_{r_{0}}Y_{pN}dr'}\\(\rho_{0}(\frac{m_{0}}{r^2c^2}
-\frac{2m^2_{0}}{r^3c^4})+X^{-1}_{pN}\left[\frac{a'}{\phi_{0}}
(1+\frac{2m_{0}}{rc^2}(1-\alpha))
+\frac{m^{2}_{0}}{r^2c^4}(1+4\alpha)\right])-X^{-1}_{pN}
\gamma^2_{\Sigma^{(e)}}\\(\frac{m_{0}}{r^2c^2}-\frac{4m^2_{0}}{r^3c^4}+\frac{2\alpha
m^2_{0}}{r^4c^4})$,
\item $p_{r0}<X^{-1}_{pN}e^{2r(\frac{m_{0}}{r^2c^2}
-\frac{2m^2_{0}}{r^3c^4})+\int^{r}_{r_{0}}Y_{pN}dr'}\int^{r}_{r_{0}}
X_{pN}e^{-2r(\frac{m_{0}}{r^2c^2}
-\frac{2m^2_{0}}{r^3c^4})-\int^{r}_{r_{0}}Y_{pN}dr'}\\(\rho_{0}(\frac{m_{0}}{r^2c^2}
-\frac{2m^2_{0}}{r^3c^4})+X^{-1}_{pN}\left[\frac{a'}{\phi_{0}}
(1+\frac{2m_{0}}{rc^2}(1-\alpha))
+\frac{m^{2}_{0}}{r^2c^4}(1+4\alpha)\right])-X^{-1}_{pN}
\gamma^2_{\Sigma^{(e)}}\\(\frac{m_{0}}{r^2c^2}-\frac{4m^2_{0}}{r^3c^4}+\frac{2\alpha
m^2_{0}}{r^4c^4}).$
\end{enumerate}
In the third case, $\Gamma>1$ leads to unstable configuration with
the following constraint
\begin{eqnarray}\nonumber
&&p_{r0}>X^{-1}_{pN}e^{2r(\frac{m_{0}}{r^2c^2}
-\frac{2m^2_{0}}{r^3c^4})+\int^{r}_{r_{0}}Y_{pN}dr'}\int^{r}_{r_{0}}
X_{pN}e^{-2r(\frac{m_{0}}{r^2c^2}
-\frac{2m^2_{0}}{r^3c^4})-\int^{r}_{r_{0}}Y_{pN}dr'}\\\nonumber
&&(\rho_{0}(\frac{m_{0}}{r^2c^2}
-\frac{2m^2_{0}}{r^3c^4})+X^{-1}_{pN}\left[\frac{a'}{\phi_{0}}
(1+\frac{2m_{0}}{rc^2}(1-\alpha))
+\frac{m^{2}_{0}}{r^2c^4}(1+4\alpha)\right])\\\nonumber
&&-X^{-1}_{pN} \gamma^2_{\Sigma^{(e)}}
(\frac{m_{0}}{r^2c^2}-\frac{4m^2_{0}}{r^3c^4}+\frac{2\alpha
m^2_{0}}{r^4c^4}),
\end{eqnarray}
where $Y_{pN}=X^{-1}_{pN}\left[\frac{a'}{\phi_{0}}
(1+\frac{2m_{0}}{rc^2}(1-\alpha))
+\frac{m^{2}_{0}}{r^2c^4}(1+4\alpha)\right]$. In this case,
$0<\Gamma<1$ is also an instability range.

\subsection{Anisotropic Fluid}

Here, we have $p_{r0}\neq p_{\phi_{0}}\neq p_{z}$ and hydrostatic
equilibrium is described by Eq.(\ref{48}).

\subsection*{Newtonian limit}

Using Eq.(\ref{a}) in (\ref{48}), we obtain condition for unstable
anisotropic filaments as
\begin{eqnarray}\label{a'}
0<\Gamma<\frac{\frac{2p_{r0}\Phi}{\phi_{0}}
+\frac{(p_{r0}-p_{\phi0})}{r}\left[\frac{c}{r}\right]'+
\rho_{0}(Z_{N}+K_{6})+K_{7}}{\left[\left[p_{r0}Z_{N}\right]'
+\frac{(p_{r0}-p_{\phi0})}{r}Z_{N}
-p_{r0}Z_{N}\frac{m_{0}}{r^2c^2}\right]},
\end{eqnarray}
where $K_{6}$ and $K_{7}$ are mentioned in Appendix \textbf{A}.
Similar to isotropic case, this implies that for
\begin{eqnarray}\nonumber
p_{r0}&\leq&r^{-1}Z^{-1}_{N}e^{(\frac{m_{0}}{r^2c^2}
-\int^{r}_{r_{0}}\frac{\Phi}{\phi}
+\frac{1}{r}\left[\frac{c}{r}\right]'dr')}
\left[\int^{r}_{r0}rZ_{N}e^{-(\frac{m_{0}}{r^2c^2}
-\int^{r}_{r_{0}}\frac{\Phi}{\phi}
+\frac{1}{r}\left[\frac{c}{r}\right]'dr')}
\left[\frac{p_{\phi_{0}}}{r}\right.\right.\\\nonumber
&+&\left.\left.\rho_{0}(1+K_{6})+K_{7}\right]dr'\right],
\end{eqnarray}
$\Gamma$ lies in $(0,1)$ and the system collapses. If
\begin{eqnarray}\nonumber
p_{r0}&>&r^{-1}Z^{-1}_{N}e^{(\frac{m_{0}}{r^2c^2}
-\int^{r}_{r_{0}}\frac{\Phi}{\phi}
+\frac{1}{r}\left[\frac{c}{r}\right]'dr')}
\left[\int^{r}_{r0}rZ_{N}e^{-(\frac{m_{0}}{r^2c^2}
-\int^{r}_{r_{0}}\frac{\Phi}{\phi}
+\frac{1}{r}\left[\frac{c}{r}\right]'dr')}
\left[\frac{p_{\phi_{0}}}{r}\right.\right.\\\nonumber
&+&\left.\left.\rho_{0}(1+K_{6})+K_{7}\right]dr'\right],
\end{eqnarray}
the system becomes unstable for $\Gamma>1$.

\subsection*{Post-Newtonian Limit}

The collapse condition of anisotropic cylindrical filaments in pN
regime is
\begin{equation}\label{b'}
\Gamma<\frac{\frac{p_{r0}}{r}U_{pN}
(\frac{m_{0}}{r^{2}c^2}-\frac{m^2_{0}}{r^3c^4})
-(p_{r0}+\rho_{0})\left[U_{pN}\right]-K_{9}}
{\left[\left[p_{r0}U_{pN}\right]'+\frac{p_{r0}}{r}U_{pN}
-\frac{p_{\phi0}}{r}V_{pN}\right]},
\end{equation}
where
\begin{eqnarray}\nonumber
U_{pN}&=&\left[b(1-\alpha\frac{m_{0}}{rc^2})
+\frac{(p_{\phi0}+\rho_{0})}{(p_{r0}+\rho_{0})}\frac{c}{r}
+\frac{1}{(p_{r0}+\rho_{0})}K_{8}\right],\\\nonumber
V_{pN}&=&\left[b(1-\alpha\frac{m_{0}}{rc^2})
\frac{(p_{r0}+\rho_{0})}{(p_{\phi0}+\rho_{0})}+\frac{c}{r}
+\frac{1}{(p_{\phi0}+\rho_{0})}K_{8}\right].
\end{eqnarray}
The values of $K_{8}$ and $K_{9}$ are given in Appendix \textbf{A}.
The system becomes to unstable for instability ranges $0<\Gamma<1$
if
\begin{itemize}
\item $p_{r0}{r}U_{pN}
(\frac{m_{0}}{r^{2}c^2}-\frac{m^2_{0}}{r^3c^4})
-(p_{r0}+\rho_{0})\left[U_{pN}\right]-K_{9}=
\left[\left[p_{r0}U_{pN}\right]'+\frac{p_{r0}}{r}U_{pN}
-\frac{p_{\phi0}}{r}V_{pN}\right]$,
\item $p_{r0}{r}U_{pN}
(\frac{m_{0}}{r^{2}c^2}-\frac{m^2_{0}}{r^3c^4})
-(p_{r0}+\rho_{0})\left[U_{pN}\right]-K_{9}
<\left[\left[p_{r0}U_{pN}\right]'+\frac{p_{r0}}{r}U_{pN}
-\frac{p_{\phi0}}{r}V_{pN}\right]$,
\end{itemize}
and becomes unstable for $\Gamma>1$ if
\begin{itemize}
\item $p_{r0}{r}U_{pN}
(\frac{m_{0}}{r^{2}c^2}-\frac{m_{0}}{r^3c^4})
-(p_{r0}+\rho_{0})\left[U_{pN}\right]-K_{9}
>\left[\left[p_{r0}U_{pN}\right]'+\frac{p_{r0}}{r}U_{pN}
-\frac{p_{\phi0}}{r}V_{pN}\right]$.
\end{itemize}

\section{Concluding Remarks}

The study of structure formation in modified gravity is an important
issue. Cosmic filamentary structures with cylindrical symmetry arise
in astrophysics both on large (cosmic web) as well as small (tidal
tails) scales. The behavior of these structures has an important
role in the formation of structures in the universe. In this paper,
we have investigated instability ranges of anisotropic cylindrically
symmetric collapsing filaments in BD theory. We have used contracted
Bianchi identities to obtain two dynamical equations of collapsing
filamentary system. By applying perturbation technique on BD as well
as dynamical equations, we separate the unperturbed (static) and
perturbed (non-static) distributions of all dynamical relations. We
have developed hydrostatic equation (collapse equation) through
perturbed configuration of second dynamical equation.

The equation of state involving adiabatic index controls the ranges
of instability for a collapsing filamentary structure. We have used
collapse equation along with equation of state to investigate the
instability ranges of both isotropic as well as anisotropic BD fluid
at Newtonian and pN limits. It is concluded that in both
approximations the adiabatic index depending upon dynamical
properties (energy density, pressure, scalar field terms and some
constraints) controls the instability ranges. We have constructed
constraints on static radial matter pressure under the effects of
scalar field. It is found that the cylindrical filamentary
structures always remain unstable for $0<\Gamma<1$, while $\Gamma>1$
is the instability range for the special case. We would like to
mention here that the instability ranges for spherical as well as
cylindrical distributions in GR depend upon $\Gamma<\frac{4}{3}$ and
$\Gamma<1$. In $f(R)$ and $f(T)$ theories, physical variables such
as density, pressure and respective modified dark terms provide the
instability ranges. The instability range of spherically symmetric
isotropic BD fluid is $\Gamma>\frac{4}{3}$ while anisotropic
spherical BD fluid always remains unstable for $0<\Gamma<1$ and
$\Gamma>1$ leads to collapse only for the special case.

\section*{Appendix A}

The non-zero components of BD equations for interior region are
\begin{eqnarray}\nonumber
&&G_{00}
=\frac{1}{\phi}(T^{m}_{00}+T^{\phi}_{00})=\frac{1}{\phi}\left({\rho}A^{2}+
\frac{\omega_{BD}}{2\phi}(\dot{\phi}^2
+\frac{A^{2}\phi^{'2}}{B^{2}})\right)\\\label{8}
&&-\frac{\dot{\phi}}{\phi}\left(\frac{2\dot{A}}{A}+\frac{\dot{B}}{B}
+\frac{\dot{C}}{C}\right)+\frac{\phi'A^2}{\phi
B^2}\left(\frac{B'}{B}+\frac{C'}{C}\right)
+\frac{A^{2}\phi''}{B^{2}\phi}-\frac{A^{2}V(\phi)}{2\phi},\\\label{9}
&&G_{01}=\frac{1}{\phi}(T^{m}_{01}+T^{\phi}_{01})
=\frac{\omega_{BD}}{\phi^2}(\dot{\phi}\phi')
+\frac{1}{\phi}\left(\dot{\phi}'
-\frac{A'\dot{\phi}}{A}-\frac{\dot{B}\phi'}{B}\right),\\\nonumber
&&G_{11}=\frac{1}{\phi}(T^{m}_{11}+T^{\phi}_{11})=
\frac{1}{\phi}\left(p_{r}B^{2}+\frac{\omega_{BD}}{2\phi}
({\phi}^{'2}+\frac{B^{2}\dot{\phi}^{2}}{A^2})\right)
+\frac{\dot{B}\ddot{\phi}}{A^{2}\phi}\\\label{10}
&&+\frac{B^2\dot{\phi}}{A^2}\left(\frac{\dot{A}}{A}
+\frac{\dot{C}}{C}\right)-\frac{\phi'}{\phi}\left(\frac{A'}{A}
+\frac{C'}{C}\right)+\frac{B^{2}V(\phi)}{2\phi},\\\nonumber
&&G_{22}=\frac{1}{\phi}(T^{m}_{22}+T^{\phi}_{22})
=\frac{1}{\phi}\left(p_{\perp}C^{2}+
\frac{\omega_{BD}}{2\phi}(\frac{\dot{C^{2}}\dot{\phi}^{2}}{A^{2}}
-\frac{C^{2}{\phi}^{'2}}{B^2})\right)+\frac{\ddot{\phi}C^2}{A^2\phi}\\\label{11}
&&+\frac{C^2\dot{\phi}}{A^{2}\phi}
\left(\frac{\dot{A}}{A}+\frac{\dot{B}}{B}\right)
-\frac{C^2\phi'}{B^{2}\phi}
\left(\frac{A'}{A}+\frac{B'}{B}\right)-\frac{C^2\phi''}{B^2\phi}
+\frac{C^{2}V(\phi)}{2\phi},\\\nonumber
&&G_{33}=\frac{1}{\phi}(T^{m}_{33}+T^{\phi}_{33})
=p_{z}+\frac{\omega_{BD}}{2\phi^2
B^2}\left[\frac{\dot{\phi}^2}{A^2}-\frac{\phi'^2}{B^2}\right]
+\frac{\ddot{\phi}}{A^2\phi}\\\label{12}
&&+\frac{\dot{\phi}}{A^2\phi}\left[\frac{\dot{A}}{A}+\frac{\dot{B}}{B}
+\frac{\dot{C}}{C}\right]
-\frac{\phi'}{B^2\phi}\left[\frac{A'}{A}+\frac{B'}{B}+\frac{C'}{C}\right]
-\frac{\phi''}{B^2\phi}+\frac{V(\phi)}{2\phi}
\end{eqnarray}
and Eq.(\ref{3}) becomes
\begin{eqnarray}\nonumber
&&\dot{\phi}\left(\frac{\dot{A}}{A}-\frac{\dot{B}}{A^2B}
-\frac{\dot{C}}{A^2B}\right)+\frac{\ddot{\phi}}{A^2}
+\phi'\left(\frac{A'}{AB^2}-\frac{B'}{B^3}-\frac{C'}{CB^2}\right)
-\frac{\phi''}{B^2}\\\label{13}
&&=\frac{1}{2\omega_{BD}+3}\left[\left(\rho+3p_{r}+p_{\phi}+p_{z}\right)
+\left(\phi\frac{dV}{d\phi}-2V\right)\right].
\end{eqnarray}

The scalar terms $K_{1}$ and $K_{2}$ of Eqs.(\ref{17}) and
(\ref{18}) are
\begin{eqnarray}\nonumber
K_{1}&=&\left(T^{\phi}_{00}\right)_{,t} A^{-1} - \left(T^{\phi}_{01}
\right)_{,r} A^{-1}B^{-2} +\left(\rho
A^{-1}+T^{\phi}_{00}A^{-2}\right)\phi^{-2}\dot{\phi}\\\nonumber
&+&T^{\phi}_{01}A^{-1}B^{-2}\phi^{2}\phi'-2T^{\phi}_{01}B^{3}A B'
-T^{\phi}_{01}A^{-2}B^{-2}A',\\\nonumber
K_{2}&=&T^{\phi}_{11}B^{-1}\phi'\phi^{-2}+\left(\rho
+T^{\phi}_{01}\right)A^{-2}B^{-1}\phi^{-2}\dot{\phi}-
\left(T^{\phi}_{01} A^{-2}B^{-2}\right)_{,t}B\\\nonumber
&-&\left(T_{11}B^{-2}\right)_{;r}B.
\end{eqnarray}

The static distribution of BD field equations is
\begin{eqnarray}\label{27}
&&\frac{\rho_0}{\phi_0}+\frac{\omega_{BD}
\phi^{'2}_0}{2B^2_{0}\phi^2_{0}}+\frac{B'_0\phi'_0}{B^{3}_0\phi_0}
+\frac{2\phi'_{0}}{B^{2}_{0}r\phi_0}+\frac{\phi''_{0}}{B^2_{0}\phi_0}
-\frac{V_0}{2\phi_0}
=\frac{1}{B_{0}^{2}r}\frac{B_0'}{B_0},\\\label{28}
&&\frac{p_{r0}}{\phi_0}+\frac{\omega_{BD}
\phi^{'2}_0}{2B^2_{0}\phi^2_{0}}
-\frac{A'_{0}\phi'_{0}}{A_{0}B^2_{0}\phi_0}
-\frac{2\phi'_{0}}{B^{2}_{0}r\phi_0}+\frac{V_0}{2\phi_0}
=\frac{1}{B_{0}^{2}r}\frac{A_0'}{A_0},\\\nonumber
&&\frac{p_{\phi0}}{\phi_0}-\frac{\omega_{BD}
\phi^{'2}_0}{2B^2_{0}\phi^2_{0}}-\frac{B'_0\phi'_0}{B^{3}_0\phi_0}
-\frac{A'_{0}\phi'_{0}}{A_{0}B^2_{0}\phi_0}
-\frac{\phi''_{0}}{B^2_{0}\phi_0} +\frac{V_0}{2\phi_0}\\\label{29}
&&=\frac{1}{B_0^2}\left[\frac{A_0''}{A_0}
+\frac{A_0'}{A_0}\frac{B_0'}{B_0}\right],\\\nonumber
&&\frac{p_{z}}{\phi_{0}}-\frac{\phi_{0}'A'_{0}}{B^{2}_{0}A_{0}\phi_{0}}
-\frac{\phi'_{0}B'_{0}}{B^{3}_{0}\phi_{0}}
-\frac{\phi'_{0}}{B^{2}_{0}r\phi_{0}}-\frac{\phi''_{0}}{B^{2}_{0}}
-\frac{\omega_{BD}\phi_{0}'^2}{2\phi_{0}^2B^{2}_{0}}
+\frac{V_{0}}{2\phi_{0}}\\\label{30}
&&=\left(\frac{A'_{0}}{r}+A''_{0}\right)\frac{1}{A_{0}B^2_{0}}
-\frac{B'_{0}}{B^{3}}\left(\frac{1}{r}+\frac{A'_{0}}{A_{0}}\right).
\end{eqnarray}
The unperturbed wave equation is
\begin{equation}\nonumber
\frac{\phi'_{0}A'_{0}}{A_{0}}-\frac{\phi'_{0}B'_{0}}{B^{2}_{0}}
+\frac{\phi'_{0}}{rB_{0}}=\frac{-1}{2\omega_{BD}+3}\left[\left(\rho_{0}
+3p_{r0}+p_{\phi0}+p_{z0}\right)+\left(\phi_{0}V_{0}-2V_{0}\right)\right].
\end{equation}
The static distribution of Eq.(\ref{17}) is identically satisfied in
static background while (\ref{18}) turns out to be
\begin{eqnarray}\label{31}
\frac{p_{r0}'}{B_{0}\phi_{0}}+
\frac{\phi'_{0}p_{r0}}{\phi^{2}_{0}B^{2}_{0}}+(\rho_0+p_{r0})\frac{A_{0'}}{A_{0}B_{0}\phi_{0}}
+\frac{1}{B_{0}\phi_{0}r}(p_{r0}-p_{\phi_{0}})-K'_{2}=0,
\end{eqnarray}
where
\begin{equation}\nonumber
K'_{2}=T^{\phi}_{11unp}\frac{\phi'_{0}}{\phi_{0}B^{2}_{0}}
-\frac{p'_{r0}}{\phi_{0}B_{0}}
-(p_{r0}+\rho_{0})\frac{A'_{0}}{A_{0}B_{0}\phi_{0}}
-\frac{(p_{r0}+p_{\phi0})}{B_{0}r\phi_{0}}
+2T^{\phi}_{11unp}\frac{B'_{0}}{B^{2}_{0}}
-\frac{(T^{\phi}_{11unp})_{;r}}{B_{0}}
\end{equation}
and the term $T^{\phi}_{11unp}$ represents unperturbed form of
energy tensor due to scalar field.

The static part of Eq.(\ref{15*}) is
\begin{eqnarray}\label{32}
p_{r0}&\overset{\Sigma^{(e)}}{=}&\frac{-V_0}{2}.
\end{eqnarray}
The perturbed form of BD field equations are
\begin{eqnarray}\nonumber
&&-\frac{2T}{B_0^2}\left[\left(\frac{c}{r}\right)''-\frac{1}{r}
\left(\frac{b}{B_0}\right)'
-\left(\frac{B_0'}{B_0}\right)\left(\frac{c}{r}\right)'
\right]=-\frac{\bar{\rho}}{\phi_{0}}-
\frac{T\rho_{0}\Phi}{\phi^{2}_{0}}\\\nonumber
&&+\frac{T\omega_{BD}\phi^{'2}_{0} b}{\phi^{2}_{0}B^{3}_{0}}
-\frac{T\omega_{BD}\bar{\Phi}\phi^{'2}_{0}}{B^3_{0}\phi^{3}_{0}}
+\frac{\omega_{BD}T\bar{\Phi}'\phi_{0}'}{B^2_{0}\phi^2_{0}}
+\frac{T\phi'_{0}}{\phi_{0}B^{2}_{0}}\left(\frac{c} {r}\right)'
+\frac{T\phi'_{0}}{\phi_{0}B^{2}_{0}}\left(\frac{\phi_{0}b}{B_0}\right)'\\\nonumber
&&-\frac{2Tb\phi'_{0}}{{\phi}B^{3}_{0}}\frac{1}{r}
+\left[\frac{T}{B^{2}_{0}r}+\frac{TB'_{0}}{B^3_{0}}\right]
\left[\frac{\Phi}{\phi_{0}}\right]'+\frac{T\Phi''}{\phi_{0}B^{2}_{0}}
-\frac{2Tb\phi''_{0}}{B_{0}^3\phi_{0}}
-\frac{T\phi'_{0}\Phi}{B^{2}_{0}\phi^{2}_{0}}\\\label{33}
&&-\frac{TV_{0}\Phi}{2\phi^{2}_{0}}-\frac{T\bar{V}}{2\phi_{0}},\\\label{34}
&&-\frac{c'}{c}+\frac{A_0'}{A_0}+\frac{b}{cB_0}=\frac{\omega_{BD}\dot{T}\Phi'\dot{\phi}}
{\phi^{2}_{0}}-\frac{\omega_{BD}\dot{T}\Phi\phi'_{0}}{\phi_{0}}
+\frac{A'_{0}\dot{T}c}{rA_{0}}-\frac{\dot{T}c'}{r}+\frac{\dot{T}b}{rB_{0}},\\\nonumber
&&-\frac{2\ddot{T}c}{rA_0^2}+
\frac{T}{B^{2}_{0}r}\left[\left(\frac{a}{A_0}\right)'+
\left(r\frac{A_0'}{A_0}\right)\left(\frac{c}{r}\right)'\right]
-\frac{2bA'_{0}}{rA_{0}B^3_{0}} =\frac{\bar{p}_{r}}{\phi_{0}}-
\frac{Tp_{r0}\Phi}{\phi^{2}_{0}}\\\nonumber
&&-\frac{T\omega_{BD}}{\phi^2_{0}} \left[\frac{\phi^{'2}_{0}
b}{B^{3}_{0}}+\Phi'-\frac{\Phi\phi^{'2}_{0}}{\phi_{0}}\right]
-\frac{T\phi'_{0}}{\phi_{0}B^{2}_{0}}
\left[\left(\frac{a}{A_{0}}\right)'+\left(\frac{c}
{r}\right)'\right]\\\nonumber
&&+\frac{2Tb\phi'_{0}}{{\phi}B^{3}_{0}}
\left[\frac{A'_{0}}{A_{0}}-\frac{1}{r}+\frac{V_{0}}{2B_{0}}\right]
-\left[\frac{T}{B^{2}_{0}r}+\frac{T
A'_{0}}{B^{2}_{0}A_{0}}-\frac{TB'_{0}}{B^3_{0}}\right]
\left[\frac{\Phi}{\phi_{0}}\right]'\\\label{35}
&&-\frac{\dot{T}b\phi'_{0}}{\phi_{0}B^{2}_{0}}
+\frac{\ddot{T}\Phi}{A^{2}_{0}\phi_{0}}
+\frac{TV_{0}\Phi}{2\phi^{2}_{0}}
+\frac{T\bar{V}}{2\phi_{0}},\\\nonumber
&&-\frac{b\ddot{T}}{A_0^2B_0}+\frac{T}{A_{0}B_0^2}
\left[\left(\frac{a}{A_0}\right)''
+\left(\frac{c}{r}\right)''+\left(\frac{2A_0'}{A_0}-\frac{B_0'}{B_0}+\frac{1}{r}\right)
\left(\frac{a}{A_0}\right)'\right.\\\nonumber
&&-\left.\left(\frac{A_0'}{A_0}+\frac{1}{r}\right)\left(\frac{b}{B_0}\right)'
+\left(\frac{A_0'}{A_0}-\frac{B_0'}{B_0}+\frac{2}{r}\right)
\left(\frac{c}{r}\right)'\right]=-\frac{\bar{p}_{\phi}}{\phi_{0}}-
\frac{Tp_{\phi0}\Phi}{\phi^{2}_{0}}\\\nonumber
&&-\frac{T\omega_{BD}\phi^{'2}_{0} b}{2\phi^{2}_{0}B^{3}_{0}}
-\frac{T\omega_{BD}\Phi'}{2\phi^{2}_{0}B^{2}_{0}}
-\frac{T\phi'_{0}}{\phi_{0}B^{2}_{0}}
\left[\left(\frac{a}{A_{0}}\right)' +\frac{T}{r^2\phi_{0}}
\left(\frac{br^2\phi'_{0}}{B_{0}}\right)'\right]\\\nonumber
&&+\frac{2Tb\phi'_{0}}{{\phi}B^{3}_{0}}\frac{A'_{0}}{A_{0}}
-\left[\frac{TB'_{0}}{B^{3}_{0}}+\frac{T
A'_{0}}{B^{3}_{0}A_{0}}\right] \left[\frac{\Phi}{\phi_{0}}\right]'
-\frac{T}{B^2_{0}\phi_{0}}\left[\frac{\Phi'}{B_{0}}\right]'
-\frac{T\Phi}{\phi_{0}B_{0}}\left[\frac{\phi'_{0}}{B_{0}}\right]'\\\label{36}
&&+\frac{2Tc}{r^2}\left[\frac{\phi'_{0}B'_{0}}{B^3_{0}}+\frac{V^2_{0}}{2}\right]
+\frac{2Tb\phi''_{0}}{B^{3}_{0}\phi_{0}}
+\frac{T\phi'_{0}\Phi}{B^{2}_{0}\phi^{2}_{0}}+\frac{TV_{0}\Phi}{2\phi^{2}_{0}}
+\frac{T\bar{V}}{2\phi_{0}},\\\nonumber
&&-T\left[\frac{2bA''_{0}}{A_{0}B^3_{0}}
+\frac{A''_{0}}{A^2_{0}B^2_{0}} +\frac{a''}{A_{0}B^2_{0}}
+\frac{A'_{0}}{B^3_{0}A_{0}}\left(\frac{b}{B_{0}}\right)'
-\frac{2bA'_{0}B'_{0}}{A_{0}B^3_{0}}
+\frac{1}{B^2_{0}}\left(\frac{a}{A_{0}}\right)'\right.\\\nonumber
&&\left.-\frac{B'_{0}}{B^3_{0}}\left(\frac{c}{r}\right)'+
\frac{2bB_{0}}{rB^3_{0}}-\frac{1}{r}\left(\frac{b}{B_{0}}\right)'
+\frac{2bA'_{0}}{A_{0}rB^3_{0}}
+\frac{A'_{0}}{A_{0}}\left(\frac{c}{r}\right)'
+\frac{1}{rB^2_{0}}\left(\frac{a}{A_{0}}\right)'\right.\\\nonumber
&&\left.+\frac{2bA'_{0}}{A_{0}B^3_{0}r}+
\frac{A'_{0}}{A_{0}B^2_{0}}\left(\frac{c}{r}\right)'
+\frac{1}{rB^2_{0}}\left(\frac{a}{A_{0}}\right)'\right]-\frac{b\ddot{T}}{A^2_{0}B_{0}}
-\frac{c\ddot{T}}{rA^2_{0}}\\\nonumber
&&=\bar{p}_{z}-\frac{T}{\phi_{0}}\left[\frac{b\phi_{0}'}{B_{0}}\right]'
-\frac{T\phi'_{0}}{\phi_{0}B^2_{0}}\left[\frac{a}{A_{0}}\right]'
-\frac{TA'_{0}}{B^2_{0}A_{0}}\left[\frac{\Phi}{\phi_{0}}\right]'
-\frac{TB'_{0}}{\phi_{0}B^3_{0}}\left[\frac{\Phi}{\phi_{0}}\right]'\\\nonumber
&&+\frac{2Tb\phi'_{0}}{B^3_{0}\phi_{0}}\left[\frac{A'_{0}}{A_{0}}
+\frac{1}{r}\right]
-\frac{\omega_{BD}T}{\phi^2_{0}B^2_{0}}\left[\frac{\Phi'\phi'_{0}}{\phi^2_{0}}-
\frac{\Phi\phi^{'2}_{0}}{\phi^3_{0}}
-\frac{\omega_{BD}b\phi^{'2}_{0}}{B^{3}_{0}\phi^2_{0}}\right]\\\label{37}
&&+\frac{\Phi''}{B_{0}\phi_{0}}
-\frac{T\phi''_{0}\Phi}{B^2_{0}\phi^2_{0}}
-\frac{TV_{0}\Phi}{\phi^2_{0}},
\end{eqnarray}
and perturbed wave equation is given by
\begin{eqnarray}\nonumber
&&\frac{\phi'_{0}}{B_{0}}\left[\frac{TaA'_{0}}{A_{0}}-Ta'
+\frac{TbB'_{0}}{B_{0}}-Tb'+\frac{Tc}{r}-Tc'\right]
+2\frac{Tb\phi''_{0}}{B_{0}}-\frac{T\Phi''}{B^{2}_{0}}\\\label{38}
&&=\frac{1}{2\omega_{BD}+3}
\times\left[\bar{\rho}+3\bar{p}_{r}+\bar{p}_{\phi}+\bar{p}_{z}+T\Phi
V_{0}-2\bar{V}\right].
\end{eqnarray}
The perturbed configuration of Eq.(\ref{15*}) is
\begin{eqnarray}\label{42}
-\bar{p}_{r}&\overset{\Sigma^{(e)}}{=}&-\frac{T\Phi
p_{r0}}{\phi_{0}}-\frac{T\Phi
V_{0}}{2\phi_{0}}+\frac{T\bar{V}}{2\phi_0}.
\end{eqnarray}
The perturbed terms $\bar{K}_{1}$ and $\bar{K}_{2}$ in
Eqs.(\ref{40}) and (\ref{41}) are described as
\begin{eqnarray}\nonumber
\bar{K_{1}}&=&\dot{T}\left[\frac{\rho_{0}\Phi}{A_{0}\phi^{2}_{0}}
+T^{\phi}_{00(p)}A^{-1}_{0}
+\left(T^{\phi}_{01(p)}\right)'A^{-1}B^{-2}-
T^{\phi}_{01(p)}A'_{0}A^{-2}B^{-2}\right.\\\nonumber
&-&\left.T^{\phi}_{01(p)}B'_{0}A^{-1}B^{-3}\right],\\\nonumber
\bar{K_{2}}&=&-T^{\phi}_{11p}\frac{\phi'_{0}}{\phi^{2}_{0}B^{2}_{0}}
+\left[-2T\Phi\phi_{0}+T\Phi'-\frac{2Tb\phi'_{0}}{\phi_{0}}\right]
\frac{T^{\phi}_{11unp}}{\phi^{2}_{0}B^{2}_{0}}
-\frac{\bar{p}'_{r}}{\phi_{0}B_{0}}\\\nonumber
&+&\frac{p'_{r0}T}{B_{0}}\left[\frac{b}{B_{0}}
+\frac{\Phi}{\phi_{0}}\right]
+\frac{(\bar{p}_{r}-\bar{p}_{\phi})}{rB_{0}\phi_{0}}+
\left[\frac{b\Phi}{B^{2}_{0}r\phi^{2}_{0}}
+\frac{1}{rB_{0}\phi^{2}_{0}}(\frac{c}{r})'\right]\\\nonumber
&\times&T(p_{r0}-p_{\phi0})
-\frac{(T^{\phi}_{10})_{,t}}{A_{0}B^{2}_{0}}
+\frac{4bB'_{0}}{B^{3}_{0}}T^{\phi}_{11unp}
-2T^{2}_{11unp}\frac{Tb'}{B^{2}_{0}}
+T^{\phi}_{11unp}\frac{Tb}{B^{2}_{0}}\\\nonumber
&-&\frac{2T^{\phi}_{11unp}B'_{0}}{B^{2}_{0}}
-\frac{(T^{\phi}_{11p})'}{B^{2}_{0}}
+\left[\frac{b}{B_{0}}+\frac{\Phi}{\phi_{0}}
+\frac{1}{B_{0}\phi_{0}} \left[\frac{a}{A_{0}}\right]'\right]
\frac{(p_{r0}+\rho_{0})A'_{0}}{A_{0}B_{0}\phi_{0}}\\\nonumber
&+&\left(\bar{\rho}+\bar{p}_{r}\right)
\frac{A'_{0}}{A_{0}B_{0}\phi_{o}}+\frac{T_{11p}B'_{0}}{B^{2}_{0}}.
\end{eqnarray}
Here $T^{\phi}_{\mu\nu(unp)}$ and $T^{\phi}_{\mu\nu(p)}$ indicate
unperturbed as well as perturbed distributions of BD energy part,
respectively.

The values of $K_{4}$ and $K_{5}$ in (\ref{58}) are given by
\begin{eqnarray}\nonumber
K_{4}&=&\rho_{0}\frac{\Phi}{\phi_{0}}+\Phi\frac{m_{0}}{r^3c^2}+
\Phi'(\frac{m_{0}}{r^2c^2}+(3+2\alpha)\frac{m^2_{0}}{r^3c^4})
+\Phi\left[1+\frac{2m_{0}}{rc^2}\right.\\\nonumber
&+&\left.\frac{m^2_{0}}{r^2c^4}-\frac{2\alpha{m_{0}}}{rc^2}
+(4\alpha+1)\frac{m^2_{0}}{r^2c^4}\right],\\\nonumber
K_{5}&=&-p_{r0}\frac{\Phi}{\phi_{0}}(1-\frac{2\alpha}{r^2c^2})
+(\rho_{0}+p_{r0})\left[\frac{a'}{\phi_{0}}
(1+\frac{2m_{0}}{rc^2}(1-\alpha))
+\frac{m^{2}_{0}}{r^2c^4}(1+4\alpha)\right]\\\nonumber
&-&\gamma^2_{\Sigma^{(e)}}(\frac{m_{0}}{r^2c^2}-\frac{4m^2_{0}}{r^3c^4}+\frac{2\alpha
m^2_{0}}{r^4c^4}).
\end{eqnarray}
The scalar field terms of (\ref{a'}) are
\begin{eqnarray}\nonumber
K_{6}&=&\gamma_{\Sigma^{(e)}}\left[\rho_{0}\Phi
+(1+\frac{\Phi'}{\phi_{0}})
+m_{0}(\frac{\Phi\phi_{0}}{r^3c^2})\right]
+\frac{\phi_{0}}{r}(1-\frac{m_{0}}{rc^2}),\\
K_{7}&=&\gamma^2_{\Sigma^{(e)}} \frac{\Phi
m^{2}_{0}\phi_{0}}{r^2c^2}-\frac{\rho_{0}a'm_{0}r}{c^2}.
\end{eqnarray}
The values of $K_{8}$ and $K_{9}$ in (\ref{b'}) are described as
\begin{eqnarray}\nonumber
K_{8}&=&\gamma_{\Sigma^{(e)}}\rho_{0}\Phi+[\frac{c}{r}]'
(1+\frac{m^{2}_{0}}{r^2c^4}-\frac{2\alpha
m_{0}}{rc^4})+(1+\frac{m^{2}_{0}}{r^2c^4}-\frac{3\alpha
m_{0}}{rc^2})\frac{\Phi'}{\phi_{0}}\\\nonumber
&+&\frac{\phi_{0}}{r}\left(1-\frac{m_{0}}{rc^2}-\frac{2\alpha
m_{0}}{r^2c^2}\right)+\Phi\phi_{0}\left(\frac{m^2_{0}}{r^2c^2}
-2\frac{m^2_{0}}{r^3c^4}\right)- \Phi\phi_{0}\frac{m^2_{0}}{r^2c^4}
-2\Phi\alpha\frac{m^2_{0}}{r^2c^4},\\\nonumber
K_{9}&=&1+\frac{\Phi'}{\phi_{0}}
+\frac{\phi_{0}}{r}\left(1-\frac{m_{0}}{rc^2}\right)
+\phi_{0}\frac{m_{0}}{r^3c^3}+\gamma^{2}_{\Sigma^{(e)}}\frac{\Phi\phi_{0}
m^2_{0}}{r^2c^2}.
\end{eqnarray}

\end{document}